\documentclass[
aps,
prb,
superscriptaddress,
amsmath, amssymb,
floatfix,
reprint,
nofootinbib,
]{revtex4-2}

\usepackage{graphicx}
\usepackage{dcolumn}
\usepackage{bm}
\usepackage[version=4]{mhchem}
\usepackage[colorlinks=True,linkcolor=blue,citecolor=blue,urlcolor=blue]{hyperref}
\usepackage{amsmath}
\usepackage{physics}
\usepackage{xspace}
\usepackage[normalem]{ulem}
\usepackage{xcolor}

\newcommand{\lafeaso}{\ce{LaFeAsO}\xspace}
\newcommand{\pba}{\ce{Pb10(PO4)6O}\xspace}
\newcommand{\pbcu}{\ce{Pb9Cu(PO4)6O}\xspace}
\newcommand{\kelvin}{\mathrm{\;K}}
\newcommand{\mev}{\mathrm{\;meV}}

\definecolor{chgblue}{RGB}{0,102,204}
\newif\ifmarked
\markedtrue
\markedfalse 
\ifmarked
  \newcommand{\chg}[1]{\textcolor{chgblue}{#1}}
  \newcommand{\chgs}[1]{\textcolor{chgblue}{\sout{#1}}}
\else
  \newcommand{\chg}[1]{#1}
  \newcommand{\chgs}[1]{}
\fi

\begin{document}


\title{Magnetic fluctuations driven by quantum geometry}
\author{Makoto Shimizu\textsuperscript{\textdaggerdbl,\S}}
\email{shimizu.makoto.es@osaka-u.ac.jp}
\affiliation{Department of Physics, Graduate School of Science, Kyoto University, Kyoto 606-8502, Japan}
\author{Chang-guen Oh\textsuperscript{\S}}
\email{cg.oh.0404@gmail.com}
\affiliation{Department of Applied Physics, The University of Tokyo, Tokyo 113-8656, Japan}
\author{Youichi Yanase}
\affiliation{Department of Physics, Graduate School of Science, Kyoto University, Kyoto 606-8502, Japan}
\date{\today}

\begin{abstract}
Using quantum distance, magnetic susceptibility in the noninteracting limit can be rigorously split into two contributions: one arises solely from band dispersion while the other stems from quantum geometric contributions. In this paper, we apply this decomposition to two materials, LaFeAsO and Pb$_9$Cu(PO$_4$)$_6$O, and demonstrate that their dominant magnetic fluctuations originate from the geometric contribution. In LaFeAsO, stripe-type antiferromagnetic fluctuations arise primarily from quantum geometry, while in Pb$_9$Cu(PO$_4$)$_6$O the geometric term suppresses antiferromagnetic fluctuations and stabilizes ferromagnetic fluctuations. Our findings highlight the essential role of quantum geometry in governing magnetic fluctuations in multiband systems, and provide a unique and quantitative framework to disentangle band-structure and wave-function-geometry effects that have often been discussed collectively as multiorbital effects.
\end{abstract}
\maketitle

\begingroup
\renewcommand{\thefootnote}{\fnsymbol{footnote}}
\setcounter{footnote}{2}
\footnotetext{Present address: Graduate School of Engineering Science, The University of Osaka, Toyonaka, Osaka 560-8531, Japan.}
\footnotetext{These authors contributed equally to this work.}
\endgroup

\section{Introduction}
In spin fluctuation theory, antiferromagnetic fluctuations are traditionally explained by Fermi surface nesting~\cite{Fedders1966,Shibatani1969,Rice1970}. In its original form, nesting refers solely to the property of energy dispersion, and the nesting vector $\bm{Q}$ is chosen to maximize the Lindhard function. This description is exact in systems without inter-band hybridization, where the irreducible susceptibility is equal to the Lindhard function. However, in multiband systems, this understanding must be extended to incorporate the overlap of Wannier functions~\cite{yanase2005,Kuroki2008,Kuroki2009,Shimizu2018}. This is because the irreducible susceptibility in multiband systems is not simply proportional to the Lindhard function and is affected by matrix elements involving wave functions. Consequently, both Fermi surface nesting and orbital overlap must be considered to understand spin fluctuations appropriately. Nesting between orbitals with negligible overlap does not enhance spin susceptibility, even if nesting appears favorable based on band dispersion alone.

The overlap of Bloch states is naturally captured by quantum geometry. In particular, the Hilbert-Schmidt quantum distance~\cite{Provost1980,shapere1989geometric}  measures the difference of Bloch wave functions. Together with the quantum‐geometric tensor~\cite{Provost1980,shapere1989geometric,ma2010abelian}, the geometric nature of Bloch states has been shown to influence a broad spectrum of electronic phenomena, such as topological properties of materials~\cite{nagaosa2010anomalous,Xiao_2010_RMP}, electromagnetic responses in superconductors~\cite{peotta2015superfluidity,torma2022superconductivity,tanaka2024,oh2025role}, anomalous Landau‐level spreading~\cite{Rhim2020,hwang2021geometric}, electronic polarizability~\cite{Resta2006,verma2024geometric}, transport properties in bilayer graphene~\cite{Oh2024Revisiting, han2026klein}, thermoelectric transport via disorder‐mediated scattering~\cite{Oh2024Thermoelectric}, optical conductivity~\cite{bhalla2022resonant,ghosh2024probing, ezawa2024analytic,ahn2022riemannian,oh2025universal,oh2025color}, and bulk--interface correspondence in singular flat‐band systems~\cite{Oh2022Bulk,Kim2023General}. 
Recently, it was shown that a necessary condition for ferromagnetic ordering can be expressed in terms of the quantum metric~\cite{Kitamura2024}, forging a deep connection between magnetism and quantum geometry~\cite{Wu2020,Kitamura2024,Heinsdorf2025,kitamura2025quantumgeometricferromagnetismsingular,kudo2025oddparitymagnetismquantumgeometry,oh2025magnetic,hu2025ferromagnetismvsantiferromagnetismnarrowband}.
By contrast, antiferromagnetic fluctuations
\chg{generally involve finite ordering vectors and therefore cannot be characterized solely by local quantities such as the quantum metric, which can dominate magnetism in the small-$q$ limit.}
\chgs{do not admit a formulation purely in the metric; instead, they emerge naturally through the quantum‐distance contribution to the non‐interacting susceptibility~\cite{Kitamura2024}.}
\chg{Instead, the relevant wave-function information is the overlap between Bloch states at $\bm{k}$ and $\bm{k}+\bm{q}$, which is quantified by the finite-$q$ quantum distance.}
This insight enables a decomposition of the irreducible susceptibility into a “band” term, dependent only on energy dispersion, and a “geometric” term, capturing wave-function‐overlap effects.
\chg{This decomposition has been applied to multiband toy models~\cite{Kitamura2024,oh2025magnetic}, clarifying the role of quantum geometry in magnetic fluctuations.}

In multiband systems, crystal structures and chemical compositions strongly shape the electronic structure and, consequently, the magnetic fluctuations. Understanding magnetism in such systems therefore requires investigations based on realistic electronic structures obtained from first-principles calculations~\cite{Mazin2008,Kuroki2008,Kuroki2009,Shimizu2018}.
While quantum geometric effects on magnetism have been clarified at a formal level, their relevance in realistic material-specific models has remained largely unexplored. In particular, it is an open question how the decomposition of the irreducible susceptibility into band and geometric contributions emerges in effective models constructed from first-principles calculations.
Addressing this issue is essential for assessing the physical significance of quantum geometry beyond idealized model studies.

In this paper, we investigate quantum geometric effects in realistic materials. Starting from first-principles calculations within density functional theory (DFT), we construct effective tight-binding models and evaluate the irreducible susceptibility, which is equivalent to the magnetic susceptibility in the non-interacting limit. We then decompose the susceptibility into the band and geometric contributions, following Refs.~\cite{Kitamura2024,oh2025magnetic}. We demonstrate that in \lafeaso and \pbcu, the dominant magnetic fluctuations arise from the geometric term. \lafeaso is a parent compound of iron-based superconductors, where stripe-type antiferromagnetic fluctuations at $\bm{Q} = (\pi, 0)$ play a key role in superconductivity~\cite{Kuroki2008,Mazin2008,Johnston2010}. While the mechanism of the antiferromagnetic fluctuations has been discussed in terms of Fermi‐surface nesting~\cite{Kuroki2008,Mazin2008}, we revisit it from the perspective of quantum geometry. The second material, \pbcu, has attracted attention for the claim of room-temperature superconductivity, but this claim has largely been refuted. 
Nevertheless, early theoretical works~\cite{Chen2024,Hirschmann2024} pointed to its intriguing quantum geometric properties, such as a large Berry curvature and the presence of Weyl points. The impact of these geometric features on the magnetic response, however, has not yet been explored.
Experimental~\cite{Guo2023,Wang2023,Zhang2024} and theoretical~\cite{Shimizu2023,SWKim2024a,SWKim2024b} studies have suggested ferromagnetic fluctuations in this material. We show that, in both materials,
the dominant magnetic fluctuations originate from the geometric term.

\section{Formalism}
We consider a tight-binding model described by the Hamiltonian
\begin{equation}
\label{eq:h0}
  H_0 = \sum_{ij} \sum_{lm} \sum_{\sigma} t_{ij}^{lm} c^{\dag}_{il\sigma} c_{jm\sigma},
\end{equation}
where $i$ and $j$ label lattice sites, $l$ and $m$ are orbital indices, and $\sigma$ denotes the spin.
In momentum space, the matrix representation of the Hamiltonian is denoted by $H_0(\bm{k})$, and its eigenvalue equation reads
$H_0(\bm{k}) \ket{u_{\lambda}(\bm{k})} = E_{\lambda}(\bm{k}) \ket{u_{\lambda}(\bm{k})}$.
Here, $E_{\lambda}(\bm{k})$ is the energy dispersion, and $\ket{u_{\lambda}(\bm{k})}$ is the corresponding Bloch wave function. The index $\lambda$ labels bands.

In this work, we focus on the static irreducible susceptibility, which we decompose into band and geometric contributions~\cite{Kitamura2024,oh2025magnetic}:
\begin{equation}
  \chi^0(\bm{q}) = \chi^0_\mathrm{band}(\bm{q}) + \chi^0_\mathrm{geom}(\bm{q}).
\end{equation}
The two terms are given explicitly as
\begin{align}
  \chi^0_{\mathrm{band}}(\bm{q}) &
  = \sum_{\bm{k}} \sum_{\lambda\lambda'} \delta_{\lambda\lambda'} F_{\lambda\lambda'}(\bm{k},\bm{q}) ,\\
  \chi^0_\mathrm{geom} (\bm{q}) & 
  = \sum_{\bm{k}} \sum_{\lambda\lambda'} \left[ 1 - d^2_{\lambda\lambda'}(\bm{k},\bm{k}+\bm{q}) - \delta_{\lambda\lambda'} \right] F_{\lambda\lambda'}(\bm{k},\bm{q}) ,
\end{align}
where $F_{\lambda\lambda'}(\bm{k},\bm{q})$ is the Lindhard function defined as
\begin{equation}
  F_{\lambda\lambda'}(\bm{k},\bm{q}) = - \frac{f(E_{\lambda}(\bm{k}))-f(E_{\lambda'}(\bm{k}+\bm{q}))}{E_{\lambda}(\bm{k})-E_{\lambda'}(\bm{k}+\bm{q})},
\label{eq:lindhard}
\end{equation}
with $f(E)$ being the Fermi-Dirac distribution function. The quantity $d_{\lambda\lambda'}(\bm{k},\bm{k}')$ represents the quantum distance between the Bloch states $\ket{u_{\lambda}(\bm{k})}$ and $\ket{u_{\lambda'}(\bm{k}')}$, given by
\begin{equation}
  d^2_{\lambda\lambda'}(\bm{k},\bm{k}') = 1 - | \langle u_{\lambda}(\bm{k}) | u_{\lambda'}(\bm{k}') \rangle |^2.
\label{eq:quantumdistance}
\end{equation}
In systems with trivial band geometry---i.e., in the absence of inter-band hybridization such that $\ev{u_\lambda(\bm{k})|u_{\lambda'}(\bm{k'})}=\delta_{\lambda\lambda'}$---the quantum distance is simplified to $d^2_{\lambda\lambda'}(\bm{k},\bm{k}') = 1 - \delta_{\lambda\lambda'}$. Consequently, the geometric contribution vanishes, and the irreducible susceptibility reduces to the band term:
$\chi^0(\bm{q}) = \chi^0_\mathrm{band}(\bm{q})$.
The detailed formalism is presented in Appendix~\ref{app:chi0}.

\chg{
In the present work, we neglect spin-orbit coupling, so that the tight-binding Hamiltonian is spin independent. The resulting susceptibility is therefore SU(2) symmetric and does not distinguish the spin direction of the fluctuation. Throughout this paper, the terms ferromagnetic and antiferromagnetic fluctuations refer to the dominant wave vector of the spin response: $\bm{q}=0$ for ferromagnetic fluctuations and finite $\bm{q}$ for antiferromagnetic fluctuations. Spin anisotropy and spin-orbit-coupling-induced modifications of the susceptibility are beyond the scope of the present analysis.
This approximation is commonly used as a first step for analyzing the wave-vector structure of magnetic fluctuations in low-energy models of iron-based superconductors~\cite{Kuroki2008,Mazin2008,Kuroki2009,Graser2009,Shimizu2018} and \pbcu~\cite{Shimizu2023,Jiang2023}.
Although \pbcu contains heavy Pb atoms, the susceptibility analyzed below is constructed within a Cu-$3d$/O-$2p$ low-energy model. Thus, spin-orbit coupling is not explicitly included in the present model.
}

\section{Antiferromagnetic fluctuations induced by quantum geometry}

\chg{
We first apply the decomposition to \lafeaso as a representative itinerant multiorbital system exhibiting stripe-type antiferromagnetic fluctuations. We note that taking into account correlation effects beyond DFT is desirable for a complete description of iron-based superconductors. Here, however, our aim is not to describe all such correlation effects, but to clarify the origin of the stripe-type magnetic response within a DFT-based itinerant low-energy model, which has been widely used as a starting point for \lafeaso and related compounds~\cite{Kuroki2008,Mazin2008,Kuroki2009,Graser2009,Shimizu2018, Johnston2010}.
}

To obtain the electronic structure of \lafeaso, we use the crystallographic parameters reported in Ref.~\cite{Kamihara2008}; the corresponding crystal structure is shown in Appendix~\ref{app:crystal}. We perform all-electron DFT calculations using the full-potential local orbital (FPLO) basis set~\cite{fplo_Koepernik1999}, the generalized gradient approximation (GGA) exchange-correlation functional~\cite{gga_Perdew1996}, and a $24 \times 24 \times 24$ $k$-mesh.

\begin{figure}[htb]
  \centering
  \includegraphics[width=\linewidth]{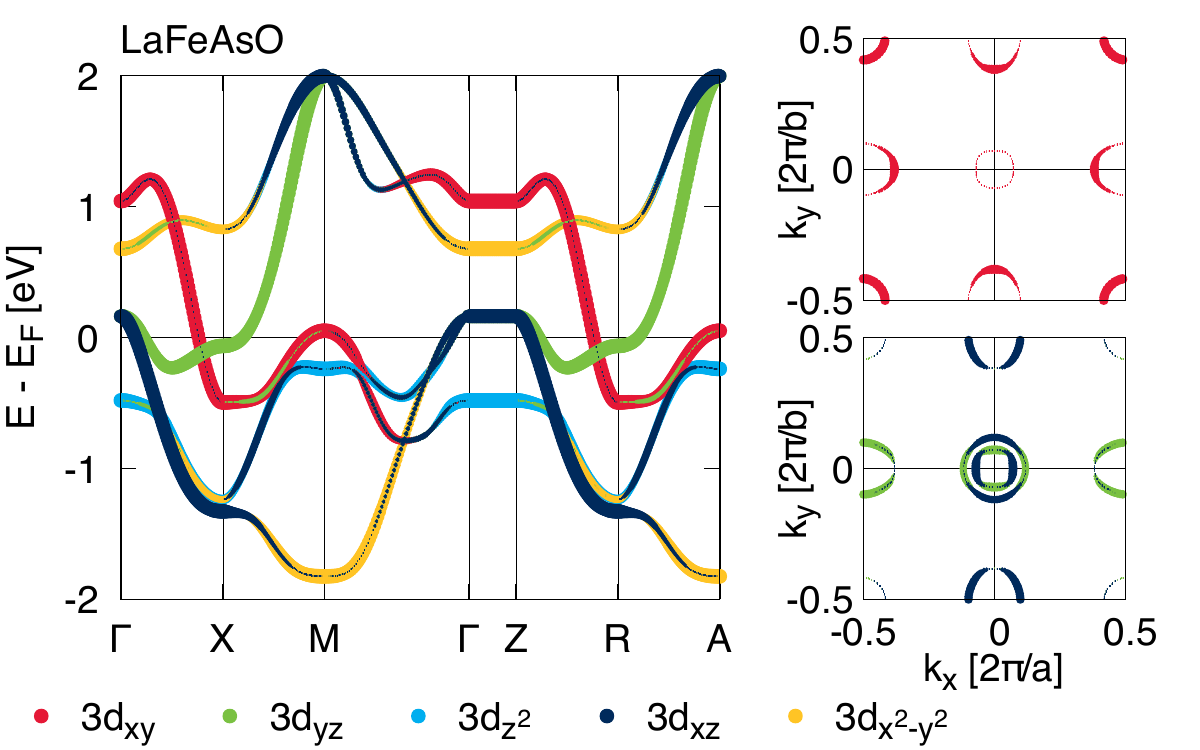}
  \caption{
    Band structure (left) and Fermi surface (right) of \lafeaso. Colors show the orbital weights of Fe $3d_{xy}$ (red), $3d_{yz}$ (green), $3d_{z^2}$ (blue), $3d_{xz}$ (navy), and $3d_{x^2-y^2}$ (yellow).
  }
  \label{fig:LaFeAsO_band_fs2d}
\end{figure}

We construct a tight-binding model comprising all ten Fe $3d$ orbitals using symmetry-conserving projected Wannier functions implemented in FPLO~\cite{wannier_Koepernik2023}. Using the glide-reflection unfolding technique~\cite{unfolding_Tomic2014}, we reduce the ten-band model to an effective five-band model.
The resulting band structure and the Fermi surface of the unfolded model are shown in Fig.~\ref{fig:LaFeAsO_band_fs2d}. Because interlayer hopping is negligibly small, the band dispersion is nearly independent of $k_z$, and the Fermi surface is almost cylindrical along the $k_z$ direction. Therefore, in the following analysis, we neglect the $k_z$ dependence of the electronic structure. Near the $\Gamma$ point [$\bm{k} = (0,0)$], two hole pockets are formed mainly by the $3d_{xz/yz}$ orbitals. Around the X point [$\bm{k} = (\frac{1}{2},0)$], an electron pocket appears with dominant contributions from the $3d_{xy}$ and $3d_{xz/yz}$ orbitals. Around the M point [$\bm{k} = (\frac{1}{2},\frac{1}{2})$], another hole pocket arises from the $3d_{xy}$ orbital. This band structure is consistent with earlier studies of \lafeaso~\cite{Kuroki2008}.

\begin{figure}[htb]
  \centering
  \includegraphics[width=\linewidth]{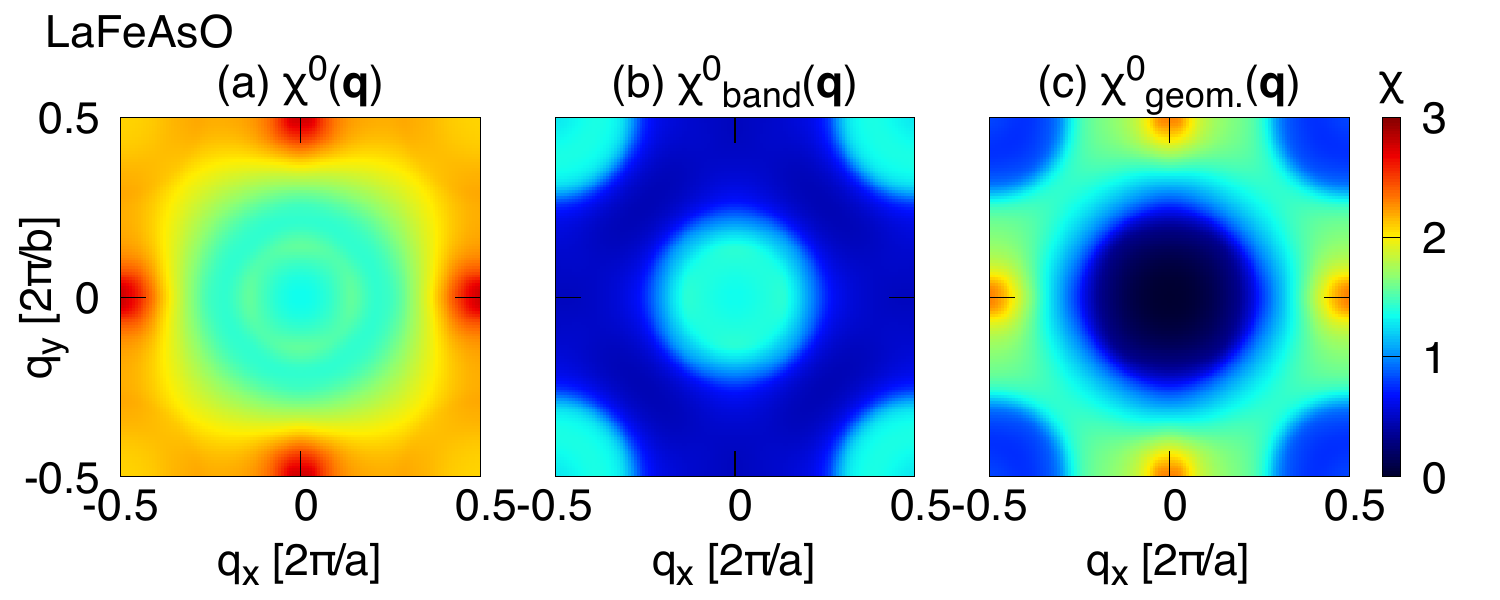}
  \caption{
    (a) The static spin susceptibility, (b) the band term and (c) the geometric term for \lafeaso at $T = 116\kelvin$.
  }
  \label{fig:LaFeAsO_chi0}
\end{figure}

We compute the static irreducible susceptibility $\chi^0(\bm{q})$, its band contribution $\chi^0_\mathrm{band}(\bm{q})$, and geometric contribution $\chi^0_\mathrm{geom}(\bm{q})$ on a $64\times64\times1$ $q$-mesh at $T = 116\kelvin$. As shown in Fig.~\ref{fig:LaFeAsO_chi0}\,(a), the total susceptibility exhibits prominent peaks at $\bm{Q} = (\frac{1}{2},0)$ and symmetry-related points.
Decomposing $\chi^0(\bm{q})$ into the band and geometric terms, we identify their respective contributions [Figs.~\ref{fig:LaFeAsO_chi0}\,(b) and \ref{fig:LaFeAsO_chi0}\,(c)]. The band contribution, $\chi^0_\text{band}(\bm q)$ [Fig. \ref{fig:LaFeAsO_chi0}\,(b)], shows only broad features and does not favor the ordering at $\bm Q$. In stark contrast, the geometric contribution, $\chi^0_\text{geom}(\bm q)$ [Fig. \ref{fig:LaFeAsO_chi0}\,(c)], exhibits dominant peaks exactly at these antiferromagnetic wave vectors. Therefore, we conclude that the stripe-type antiferromagnetic instability in LaFeAsO is decisively driven by the quantum geometry of the Bloch states.

\section{Ferromagnetic fluctuations induced by quantum geometry}

\chg{We next consider electron-doped \pbcu within a Cu-$d$/O-$p$ itinerant model. We emphasize that the structural model, metallicity, and correlation effects in this material remain under debate. Therefore, our results should not be interpreted as a definitive statement on the ground state of \pbcu. Rather, our purpose is to analyze, within the doped itinerant model proposed in Ref.~\cite{Shimizu2023}, whether the leading ferromagnetic response originates from the band contribution or from the geometric contribution.}

We construct the crystal structure of \pbcu following the method in Ref.~\cite{Shimizu2023}. The process starts from the structure of \pba determined by Krivovichev and Burns~\cite{Krivovichev1992}, which belongs to the $P63/m$ (No.~176) space group and has a partially occupied (one-quarter) O(4) site. To simplify the model, we select one of four symmetry-equivalent O(4) sites, thereby reducing the symmetry to $P3$ (No.~143). 
According to Ref.~\cite{Si2023}, substitution of Cu at the Pb(2) site is energetically favorable. We thus replace one of the four Pb atoms with a Cu atom and relax the internal atomic positions. The crystal structure is shown in Appendix~\ref{app:crystal}. Using experimentally determined lattice constants~\cite{Lee2023a}, we perform all-electron DFT calculations with the FPLO basis set~\cite{fplo_Koepernik1999}, the GGA exchange-correlation functional~\cite{gga_Perdew1996}, and a $12 \times 12 \times 12$ $k$-mesh.

\begin{figure}[htb]
  \centering
  \includegraphics[width=\linewidth]{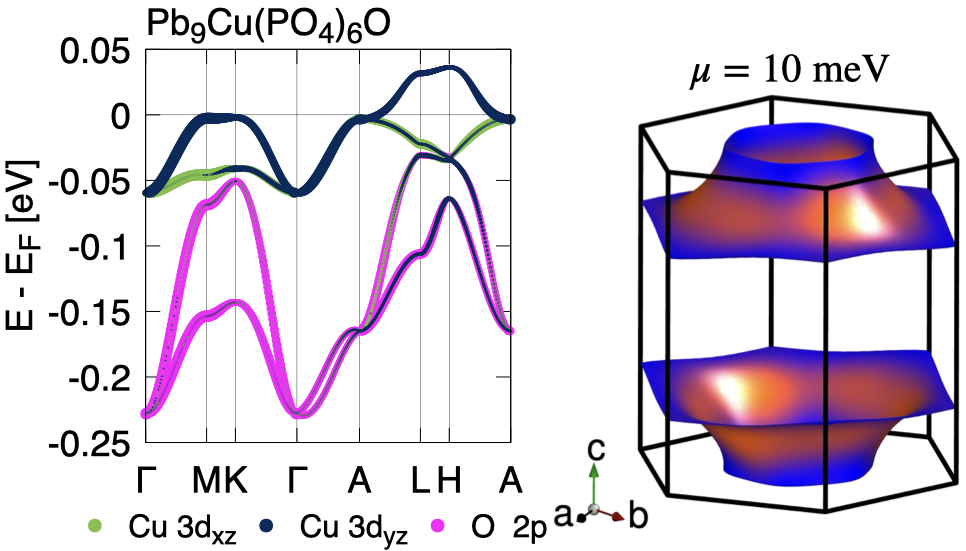}
  \caption{
    Band structure (left) and Fermi surface (right) of \pbcu. The Fermi surface is displayed for the chemical potential, $\mu = 10 \mev$. Orange (blue) indicates high (low) Fermi velocity as determined by the FPLO calculation.
  }
  \label{fig:pbcu_band_fs}
\end{figure}

A four-band tight-binding model is constructed from projected Wannier functions~\cite{wannier_Koepernik2023}, incorporating the Cu $3d_{xz}$, $3d_{yz}$ and O $2p_x$, $2p_y$ orbitals. The resulting band structure and Fermi surface are shown in Fig.~\ref{fig:pbcu_band_fs}. For stoichiometric {\pbcu}, the total electron filling is $n = 7$. To explore ferromagnetism in the electron-doped region, as suggested by the fluctuation exchange approximation~\cite{Shimizu2023}, we set the chemical potential to $\mu = 10\mev$, corresponding to an electron doping level of $\Delta n \approx 0.4$.

We calculate the static spin susceptibility $\chi^0$, its band part $\chi^0_\mathrm{band}$, and geometric part $\chi^0_\mathrm{geom}$ on a $64\times64\times32$ $k$-mesh at $T = 116\kelvin$, considering only the spin of Cu $3d$ orbitals.
\chg{
Specifically, we diagonalize the full Cu-$3d$/O-$2p$ Hamiltonian, while the orbital indices entering the spin susceptibility are restricted to the Cu $3d$ orbitals.
Thus, for this Cu-$3d$-projected susceptibility, the Bloch-state overlap appearing in the full-space definition is replaced by its projected counterpart, $\left| \bra{u_{\lambda}(\bm{k})} P_{3d} \ket{u_{\lambda'}(\bm{k}+\bm{q})} \right|^2$, where $P_{3d}$ is the projector onto the Cu $3d$ subspace. 
Since $P_{3d}$ is a Hermitian projector, this overlap factorizes as 
$\left| \bra{u_{\lambda}(\bm{k})} P_{3d} \ket{u_{\lambda'}(\bm{k}+\bm{q})} \right|^2 = w_{\lambda}(\bm{k})\, w_{\lambda'}(\bm{k}+\bm{q}) \left[ 1 - \tilde{d}^{\,2}_{\lambda\lambda'}(\bm{k},\bm{k}+\bm{q}) \right]$, 
where $w_{\lambda}(\bm{k}) = \bra{u_{\lambda}(\bm{k})} P_{3d} \ket{u_{\lambda}(\bm{k})}$ is the Cu-$3d$ orbital weight of the Bloch state, and $\tilde{d}^{\,2}_{\lambda\lambda'}$ is the Hilbert--Schmidt quantum distance between the normalized projected states $\ket{\tilde{u}_{\lambda}(\bm{k})} = P_{3d}\ket{u_{\lambda}(\bm{k})}/\sqrt{w_{\lambda}(\bm{k})}$, defined within the Cu $3d$ subspace. Hence, the interpretation of the geometric contribution based on the quantum distance persists after the projection, now supplemented by the orbital-weight factor $w_{\lambda}w_{\lambda'}$, which encodes the orbital matrix-element effect. Because $w_{\lambda}(\bm{k}) < 1$, the projected overlap does not reduce to $\delta_{\lambda\lambda'}$ at $\bm{q}=0$, and the Cu-$3d$-projected geometric contribution does not vanish there; its diagonal part reduces to $w_{\lambda}(\bm{k})^2 - 1 \le 0$, quantifying the spectral-weight leakage into the O-$2p$ orbitals.
}

\begin{figure}[htb]
  \centering
  \includegraphics[width=\linewidth]{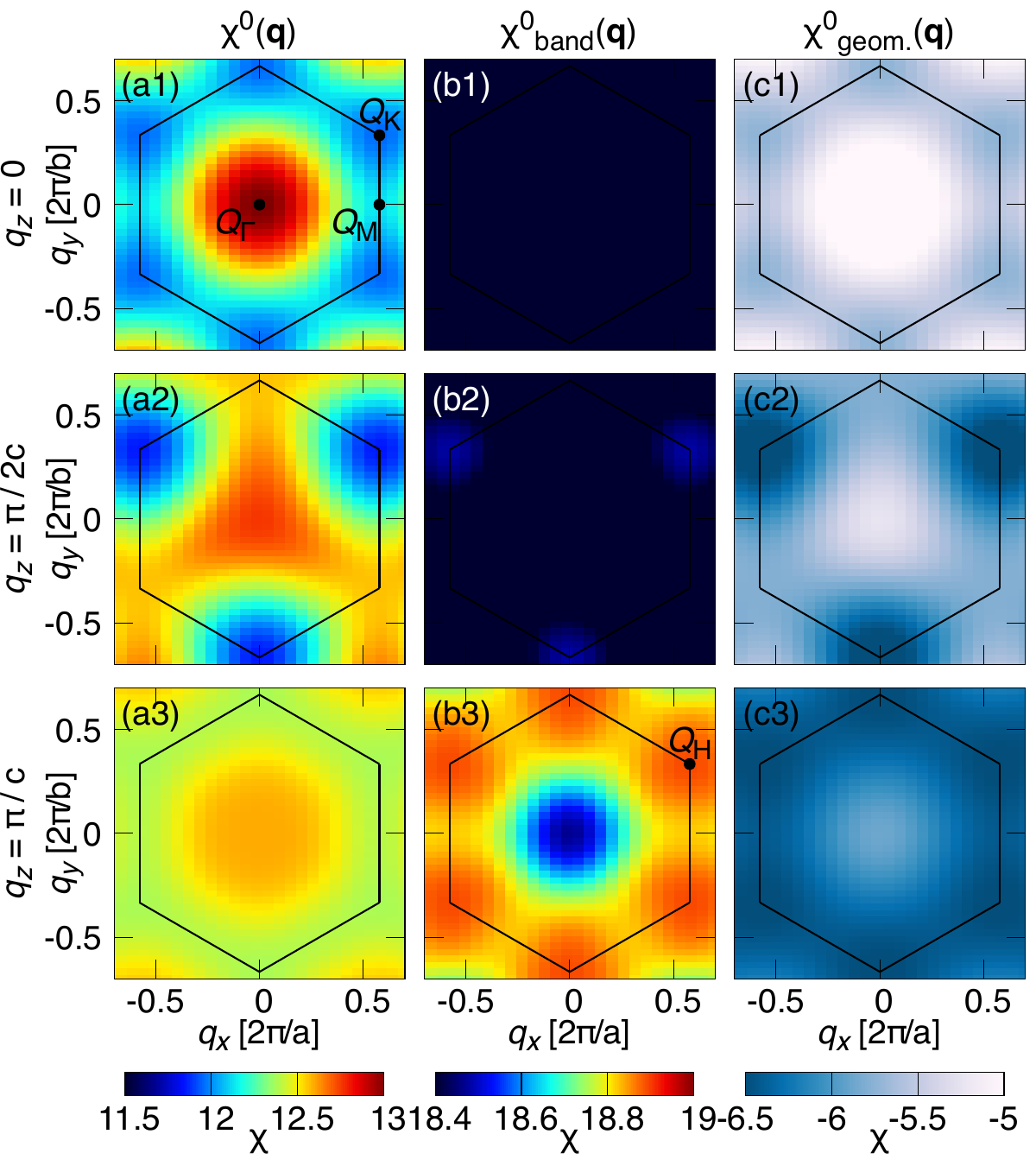}
  \caption{
    Static irreducible susceptibility of electron-doped \pbcu ($\mu = 10\mev$) at $T = 116\kelvin$. Total susceptibility $\chi^0$ (left), the band term $\chi^0_\mathrm{band}$ (middle) and the geometric term $\chi^0_\mathrm{geom}$ (right) are plotted. Cuts through $q_z = 0$ (top), $q_z = \pi/2c$ (middle) and $q_z = \pi/c$ (bottom) are shown.
  }
  \label{fig:pbcu_chi0}
\end{figure}

The total static susceptibility [Figs.~\ref{fig:pbcu_chi0}\,(a1)--(a3)] exhibits a peak at $\bm{Q}_{\Gamma} = (0,0,0)$, indicating a ferromagnetic fluctuation. Upon decomposing $\chi^0$, we find that the band term [Figs.~\ref{fig:pbcu_chi0}\,(b1)--(b3)] shows peaks at $\bm{Q}_{H} = (\frac{1}{2},\frac{1}{2},\frac{1}{2})$, which would favor antiferromagnetic fluctuations if the geometric term were absent. However, the geometric contribution, $\chi^0_{\text{geom}}(\bm q)$ [Figs.~\ref{fig:pbcu_chi0}\,(c1)--(c3)], is negative for all $\bm{q}$, with particularly large magnitude at $\bm{Q}_{H}$, thereby suppressing the antiferromagnetic tendency. As a result, the total susceptibility $\chi^0(\bm{q})$ at $\bm{Q}_H$ is diminished, leaving the ferromagnetic fluctuation at $\bm{Q}_\Gamma$ as the leading instability.
\chgs{We thus conclude that quantum geometry stabilizes ferromagnetism in electron-doped \pbcu by actively suppressing the competing antiferromagnetic order favored by the band structure.}
\chg{We thus conclude that, within the present doped itinerant model, the Cu-$3d$-projected geometric contribution stabilizes the ferromagnetic response by suppressing the competing antiferromagnetic tendency favored by the band contribution.}

\begin{figure*}[htb]
  \centering
  \includegraphics[width=\linewidth]{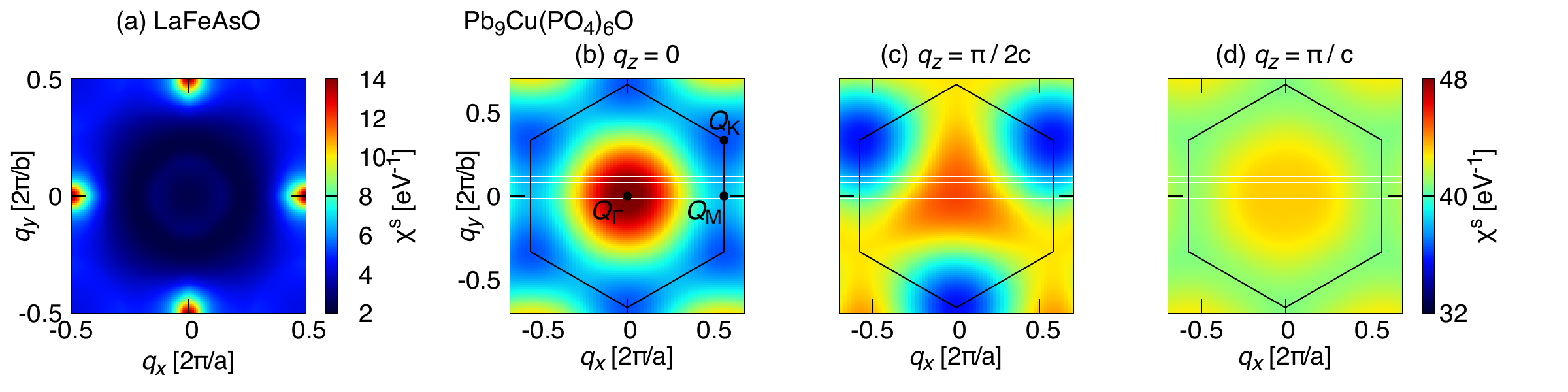}
  \caption{
      Static spin susceptibility $\chi^\mathrm{s}(\bm{q})$ calculated within RPA.
      (a) LaFeAsO for onsite Coulomb interaction parameters $U = 0.8\mathrm{\;eV}$, $V = U/2$, and $J = J' = U/4$ at $T = 116\kelvin$, showing a pronounced peak at $\bm{Q} = (\frac{1}{2}, 0)$.
      (b-d) \pbcu ($\mu = 10\mev$) for $U = 100\mev$, $V = U/2$, and $J = J' = U/4$ at $T = 116\kelvin$, showing $\chi^\mathrm{s}(\bm{q})$ for $q_z =0$, $\pi/2c$, and $\pi/c$, respectively. A ferromagnetic peak at $\bm{Q} = (0, 0, 0)$ is enhanced, while antiferromagnetic fluctuations are suppressed.
  }
  \label{fig:rpa_chis}
\end{figure*}

\section{Discussion}

\chgs{We have demonstrated that magnetic fluctuations in two realistic materials are predominantly governed by quantum geometry, but through strikingly different mechanisms.}
\chg{We have shown that the band-geometric decomposition provides a useful way to analyze the microscopic origin of magnetic fluctuations in first-principles-based multiorbital models. In the two systems studied here, the dominant magnetic responses are strongly affected by the geometric contribution, but through distinct mechanisms.}
This distinction originates from the sign of the geometric contribution, $\chi^0_{\text{geom}}$: In \lafeaso, it is positive and enhances antiferromagnetic fluctuations at $\bm{Q} = (\frac{1}{2},0)$. In contrast, in electron-doped \pbcu, it is negative and suppresses the antiferromagnetic fluctuations favored by the band structure, allowing ferromagnetic fluctuations at $\bm{Q} = (0, 0, 0)$ to become dominant.

A crucial question is whether this geometrically-driven magnetic tendency persists in the presence of electron-electron interactions. To address this, we employ the random phase approximation (RPA) and calculate spin susceptibility in interacting systems; the RPA formalism is summarized in Appendix~\ref{app:rpa}. As confirmed by our results (Fig.~\ref{fig:rpa_chis}), the peak positions of the magnetic susceptibility remain unchanged upon including electron-electron interactions. In \lafeaso, the peak at $\bm{Q} = (\frac{1}{2}, 0)$ is enhanced by interactions, consistent with previous studies~\cite{Kuroki2008}. In \pbcu, the ferromagnetic peak at $\bm{Q} = (0, 0, 0)$ is likewise intensified, in agreement with Ref.~\cite{Shimizu2023}. These results indicate that the geometric origin of magnetic fluctuations persists at least in the weak-coupling regime.

\chgs{The geometric origin of the antiferromagnetic fluctuations observed in \lafeaso may be common in other iron-based superconductors, which share similar Fermi surfaces and orbital characters.}
\chg{
The geometric origin of the antiferromagnetic fluctuations observed in \lafeaso may have implications for superconductivity in iron-based superconductors. In \lafeaso and related compounds, stripe-type antiferromagnetic fluctuations at wave vectors such as $\bm{Q} = (\frac{1}{2}, 0)$ have been widely discussed as a possible source of spin-fluctuation-mediated pairing~\cite{Kuroki2008,Kuroki2009}. Our calculation does not solve the superconducting pairing problem and therefore does not by itself determine the superconducting state. Nevertheless, it shows that the magnetic fluctuation commonly invoked as the pairing glue is not purely a consequence of Fermi-surface nesting or band dispersion. Rather, within the present DFT-based itinerant model, it is strongly enhanced by the geometric contribution associated with Bloch-state overlaps. Thus, the present result provides a microscopic classification of the source of the magnetic fluctuation relevant to pairing.
}

Quantum geometry thus emerges as a key factor in controlling magnetic responses across the family of iron-based superconductors. For example, some compounds in this class display other types of antiferromagnetic fluctuations with $\bm{Q} = (\frac{1}{2},\frac{1}{2})$~\cite{Thomale2011} and $\bm{Q} = (\frac{1}{4}, \frac{1}{4})$~\cite{Ma2009,Lucking2018} in addition to the stripe-type fluctuations at $\bm{Q} = (\frac{1}{2}, 0)$. The interplay between these magnetic fluctuations gives rise to a variety of superconducting gap symmetries, including $d_{x^2-y^2}$-wave, nodal $s_{\pm}$-wave, and nodeless $s_{\pm}$-wave states~\cite{Kuroki2009}.  Applying the band-geometry decomposition to this broader class of materials could elucidate how wave-function geometry governs the competition between distinct magnetic instabilities and the resulting superconducting gap structures.

In the case of \pbcu, the ferromagnetic fluctuation is revealed to originate from quantum geometry. While Ref.~\cite{Shimizu2023} proposed that ferromagnetism might arise from the energy dispersion along the $k_z$ direction, our analysis indicates that the dispersion alone is insufficient.  The geometric term is indispensable for stabilizing the ferromagnetic response.

In summary, we have demonstrated that quantum geometry, quantified via the quantum distance, plays a decisive role in driving both antiferromagnetic and ferromagnetic fluctuations in realistic materials. By applying the band–geometric decomposition of susceptibility to first-principles-based tight-binding models, we have revealed that the dominant magnetic fluctuations in \lafeaso and \pbcu originate from the geometric contribution. Our results highlight the crucial importance of quantum geometry in understanding magnetism in \chgs{multiband systems}\chg{realistic materials} and suggest that this framework can be broadly extended to other correlated electron systems.

The present framework opens a route to systematically classify magnetic fluctuations in multiband systems by disentangling band-dispersion and wave-function-geometry contributions in a unique and quantitative manner. While the importance of orbital matrix elements beyond simple Fermi-surface nesting has been recognized in various contexts, the band–geometric decomposition based on the quantum distance provides a unified and quantitative framework to organize such effects.
Applying this approach to a broad range of materials will facilitate the identification of compounds in which magnetism is predominantly controlled by wave-function geometry, thereby providing a practical guideline for exploring magnetic and superconducting materials.

M.S. is supported by ISHIZUE 2025 of Kyoto University.
Y.Y. is supported by JSPS KAKENHI (Grants No.~JP22H01181, No.~JP22H04933, No.~JP23K17353, No.~JP23K22452, No.~JP24K21530, No.~JP24H00007, No.~JP25H01249, and No.~JP26H02016).
C.O. was supported by JSPS KAKENHI Grant No.~JP25KF0186.
The computations in this work were performed using the facilities of the Supercomputer Center, the Institute for Solid State Physics, the University of Tokyo (ISSPkyodo-SC-2025-Ca-0062).

\appendix

\section{Irreducible susceptibility with quantum distance}
\label{app:chi0}
In this section, we write the components of the eigenvectors as
\begin{equation}
  \ket{u_{\lambda}(\bm{k})} =
  \begin{pmatrix}
    a_{\lambda}^{1}(\bm{k}) \\
    a_{\lambda}^{2}(\bm{k}) \\
    a_{\lambda}^{3}(\bm{k}) \\
    \vdots
  \end{pmatrix},
\end{equation}
where $l$ labels internal degrees of freedom such as orbitals and sublattices.
%
The irreducible susceptibility is obtained as
\begin{equation}
\begin{split}
  & \chi^0_{ll';mm'}(\bm{q},i\nu_m) \\
  &= - \sum_{\bm{k}} \frac{1}{\beta}\sum_{n} G^0_{l'm}(\bm{k},i\omega_n) G^0_{m'l}(\bm{k}+\bm{q},i\omega_n+i\nu_m) \\
  &= - \sum_{\bm{k}} \frac{1}{\beta}\sum_{n} \sum_{\lambda\lambda'} \frac{a_{\lambda}^{l'*}(\bm{k})a_{\lambda}^{m}(\bm{k})}{i\omega_n-E_{\lambda}(\bm{k})} \frac{a_{\lambda'}^{m'*}(\bm{k}+\bm{q})a_{\lambda'}^{l}(\bm{k}+\bm{q})}{i\omega_n+i\nu_{m}-E_{\lambda'}(\bm{k}+\bm{q})} \\
  &= -\sum_{\bm{k}} \sum_{\lambda,\lambda'} a_{\lambda}^{l'*}(\bm{k})a_{\lambda}^{m}(\bm{k}) a_{\lambda'}^{m'*}(\bm{k}+\bm{q})a_{\lambda'}^{l}(\bm{k}+\bm{q}) \\
  & \qquad\qquad\qquad\quad
  \times \frac{f(E_{\lambda}(\bm{k}))-f(E_{\lambda'}(\bm{k}+\bm{q}))}{E_{\lambda}(\bm{k})-E_{\lambda'}(\bm{k}+\bm{q})+i\nu_m}.
\end{split}
\end{equation}
Note that $E_{\lambda}(\bm{k})$ includes the chemical potential.
For $\nu_{m}= 0$, the irreducible susceptibility becomes
\begin{equation}
\begin{split}
  &\chi^0_{ll';mm'}(\bm{q},i\nu_0) \\
  &= \sum_{\bm{k}} \sum_{\lambda\lambda'} a_{\lambda}^{l'*}(\bm{k})a_{\lambda}^{m}(\bm{k}) a_{\lambda'}^{m'*}(\bm{k}+\bm{q})a_{\lambda'}^{l}(\bm{k}+\bm{q}) F_{\lambda\lambda'}(\bm{k},\bm{q}),
\end{split}
\end{equation}
where $F_{\lambda\lambda'}(\bm{k},\bm{q})$ is the Lindhard function defined in Eq.~\eqref{eq:lindhard}.
The static susceptibility is defined as
\begin{equation}
\begin{split}
  &\chi^{0}(\bm{q}) \\
  &= \sum_{lm} \chi^0_{ll;mm}(\bm{q},i\nu_0) \\
  &= \sum_{\bm{k}} \sum_{\lambda\lambda'} \sum_{lm} a_{\lambda}^{l*}(\bm{k})a_{\lambda}^{m}(\bm{k}) a_{\lambda'}^{m*}(\bm{k}+\bm{q})a_{\lambda'}^{l}(\bm{k}+\bm{q}) F_{\lambda\lambda'}(\bm{k},\bm{q}).
\end{split}
\end{equation}
It can be written as
\begin{equation}
  \chi^{0}(\bm{q})
  = \sum_{\bm{k}} \sum_{\lambda\lambda'} \left[ 1 - d^2_{\lambda\lambda'}(\bm{k},\bm{k}+\bm{q}) \right] F_{\lambda\lambda'}(\bm{k},\bm{q}),
\end{equation}
where $d^2_{\lambda\lambda'}$ is the quantum distance defined in Eq.~\eqref{eq:quantumdistance}.

In this study, we investigate the static susceptibility in terms of its contributions from band dispersions and quantum geometry.
\begin{equation}
  \chi^0(\bm{q}) = \chi^0_\mathrm{band}(\bm{q}) + \chi^0_\mathrm{geom}(\bm{q}).
\end{equation}
In systems with trivial geometry, where there is no hybridization between different bands, $d^2_{\lambda\lambda'}(\bm{k},\bm{k}') = 1 - \delta_{\lambda\lambda'}$. Therefore, $\chi^0(\bm{q})$ depends only on the band dispersion as \begin{equation}
  \chi^0_{\mathrm{band}}(\bm{q})
  = \sum_{\bm{k}} \sum_{\lambda\lambda'} \delta_{\lambda\lambda'} F_{\lambda\lambda'}(\bm{k},\bm{q}) .
\end{equation}
We call this contribution a band term. The remaining contribution to the static susceptibility reflects the quantum geometry of the system.
\begin{equation}
  \chi^0_\mathrm{geom} (\bm{q}) 
  = \sum_{\bm{k}} \sum_{\lambda\lambda'} \left[ 1 - d^2_{\lambda\lambda'}(\bm{k},\bm{k}+\bm{q}) - \delta_{\lambda\lambda'} \right] F_{\lambda\lambda'}(\bm{k},\bm{q}).
\end{equation}
Again, the geometric term becomes zero for systems with trivial geometry.

\section{Crystal structures}
\label{app:crystal}
\begin{figure}[h]
    \centering
    \includegraphics[width=0.9\linewidth]{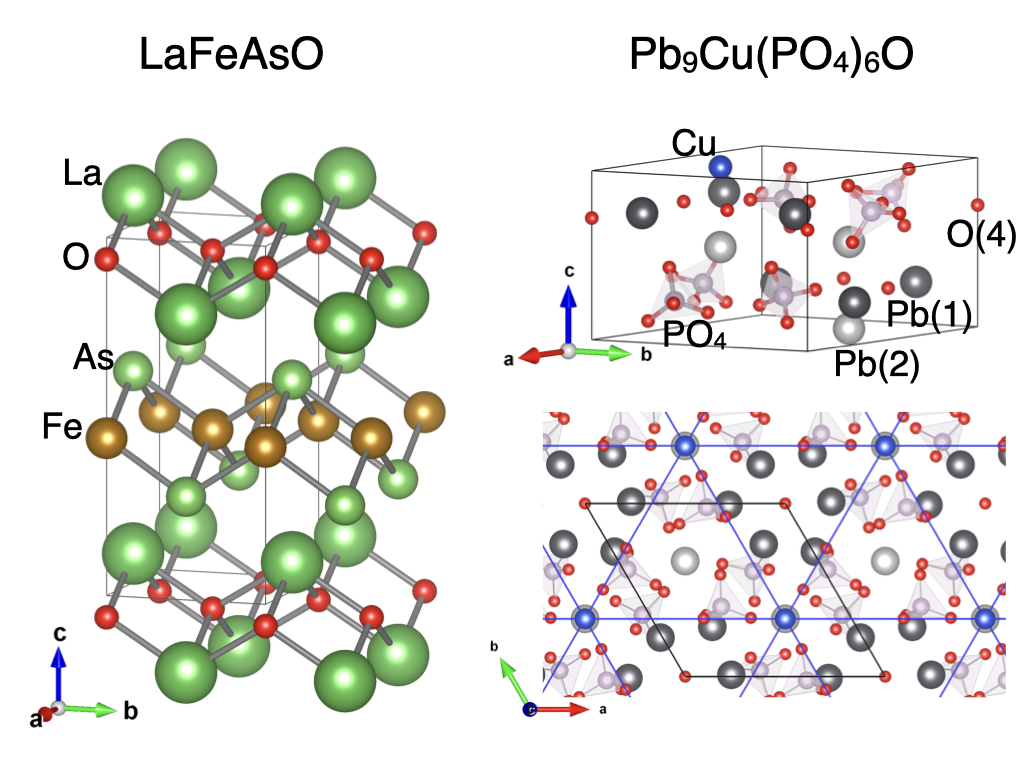}
    \caption{
      Crystal structures of LaFeAsO (left) and \pbcu (right).
    }
    \label{fig:crystal}
\end{figure}

In Fig.~\ref{fig:crystal}, we show the crystal structures used for the DFT calculations in the main text.

\section{Random Phase Approximation}
\label{app:rpa}
We consider an extended Hubbard model, whose Hamiltonian is given by
\begin{equation}
  H = H_0 + H_\mathrm{int},
\end{equation}
where $H_0$ is the tight-binding Hamiltonian~\eqref{eq:h0}, and $H_\mathrm{int}$ represents the two-particle interactions given by
\begin{equation}
\label{eq:h_hubbard}
\begin{split}
  H_\mathrm{int}
  =  \sum_{i} \biggl[
     &\frac{U}{2} \sum_{l}\sum_{\sigma} n_{il\sigma} n_{il\bar{\sigma}} 
   + \frac{V}{2} \sum_{l{\neq}m}\sum_{\sigma\sigma'} n_{il\sigma} n_{im\sigma'} \\
   &- \frac{J}{2} \sum_{l{\neq}m} \bm{S}_{il} \cdot \bm{S}_{im}
   + \frac{J'}{2} \sum_{l{\neq}m} c^\dag_{il\sigma} c^\dag_{il\bar{\sigma}} c_{im\bar{\sigma}} c_{im\sigma}
   \biggr].
\end{split}
\end{equation}
Note that we follow the notation in Ref.~\cite{Graser2009} and that this definition is different from Refs.~\cite{Kubo2007,Kuroki2008}. The parameters $(U,V,J,J')$ in this notation have the relations,
\begin{equation}
  \tilde{U} = U, \quad
  \tilde{U}' = V + \frac{J}{4}, \quad
  \tilde{J} = \frac{J}{2}, \quad
  \tilde{J}' = J',
\end{equation}
where $(\tilde{U},\tilde{U}',\tilde{J},\tilde{J}')$ are the parameters defined in Refs.~\cite{Kubo2007,Kuroki2008}.

We apply the random phase approximation (RPA) to the Hubbard model. Within RPA, the spin susceptibility is obtained by solving the following equation:
\begin{equation}
\begin{split}
  &\chi^\mathrm{s}_{ll',mm'}(q) \\
  &= \chi^0_{ll',mm'}(q)
  - \sum_{nn'oo'}\chi^0_{ll',nn'}(q) U^\mathrm{s}_{nn',oo'} \chi^\mathrm{s}_{oo',mm'}(q),
\end{split}
\end{equation}
where $U^\mathrm{s}$ represents the bare interactions in the spin channel. Components of the bare interactions are
\begin{equation}
\begin{split}
  &U^\mathrm{s}_{ll,ll} = U,\quad
  U^\mathrm{s}_{ll,mm} = \cfrac{J}{2},\quad
  \\
  &U^\mathrm{s}_{lm,lm} = V + \cfrac{J}{4},\quad
  U^\mathrm{s}_{lm,ml} = J'.
\end{split}
\end{equation}
In this study, we evaluate the static spin susceptibility,
\begin{equation}
  \chi^\mathrm{s}(\bm{q}) = \sum_{lm} \chi^\mathrm{s}_{ll,mm}(\bm{q},i\nu_0),
\end{equation}
for $\nu_0 = 0$.

%

%
%

\bibliography{quantumgeometry}

\end{document}